\documentclass[prd,nofootinbib,floats,superscriptaddress,eqsecnum,tightenlines,11pt]{revtex4}
\usepackage{hyperref}
\usepackage{graphicx}
\usepackage{amsmath,amssymb,amsfonts,amsthm,latexsym,stmaryrd}
\usepackage{marginnote}
\usepackage{color}

\newcommand{\no}{\noindent}
\newcommand{\nb}{\nonumber}
\def\beq{\begin{equation}}
\def\eeq{\end{equation}}
\newcommand{\bea}{\begin{eqnarray}}
\newcommand{\eea}{\end{eqnarray}}

\newcommand\Ost{Ostrogradsky }

\usepackage{ulem}
\usepackage{soul}

\def\m{\mu}
\def\n{\nu}
\def\r{\rho}
\def\s{\sigma}
\def\a{\alpha}
\def\b{\beta}
\def\g{\gamma}
\def\d{\delta}
\def\l{\ell}
\def\de{\partial}
\def\e{\epsilon}

\begin{document}

\title{Beyond Lovelock gravity: Higher derivative metric theories}
\author{M. Crisostomi}
\affiliation{Institute of Cosmology and Gravitation, University of Portsmouth, Portsmouth, PO1 3FX, UK}
\author{K. Noui}
\affiliation{Laboratoire de Math\'ematiques et Physique Th\'eorique, CNRS, Universit\'e Fran\c cois Rabelais, Parc de Grandmont, 37200 Tours, France}
\affiliation{Laboratoire Astroparticule et Cosmologie, CNRS, Universit\'e Paris Diderot Paris 7, 75013 Paris, France}
\author{C. Charmousis}
\affiliation{Laboratoire de Physique  Th\'eorique, CNRS, Universit\'e  Paris-Sud, 91405 Orsay, France}
\author{D.  Langlois}
\affiliation{Laboratoire Astroparticule et Cosmologie, CNRS, Universit\'e Paris Diderot Paris 7, 75013 Paris, France}

\date{\today}

\begin{abstract}
\no We consider theories describing the dynamics of  a four-dimensional metric, 
 whose Lagrangian is  diffeomorphism invariant and depends at most on second derivatives of the metric.
Imposing   degeneracy conditions we find a set of Lagrangians that,
apart form the Einstein-Hilbert one, are either trivial or contain more than two degrees of freedom. 
Among the partially degenerate theories, we recover Chern-Simons gravity, endowed with  constraints whose structure suggests the presence of instabilities.
Then, we enlarge the class of parity violating theories of gravity by introducing new ``chiral scalar-tensor theories''. Although they all raise the same concern as Chern-Simons gravity, they can nevertheless make sense as low energy effective field theories or, by restricting them to the unitary gauge (where the scalar field is uniform), as Lorentz breaking theories with a parity violating sector.
\end{abstract}

\maketitle

\section{Introduction}
\no It is well known that the Einstein-Hilbert plus cosmological constant action is the unique diffeomorphism (diff) invariant action for a  four dimensional  metric, whose equations of motion (EOM) are at most of second order \cite{Lovelock:1971yv}.
The metric field contains only two physical degrees of freedom, corresponding to a massless spin-2 field.
Any other action leads to higher order EOM (or trivial ones).

According to Ostrogradsky's analysis \cite{Ostrogradski, Woodard:2006nt} higher order EOM may signal, under certain hypotheses, the presence of instabilities which generically render the theory  pathological. 
However, recent examples of theories (breaking the above hypotheses) show that having higher order EOM is not equivalent to having ghost(s) propagating in the theory. 
In other words, although it is clear that the presence of an \Ost mode necessarily implies (by definition) 
higher order Euler-Lagrange equations, the reverse is not true.

A prime example is the one of scalar-tensor theories beyond Horndeski which were 
introduced in \cite{Zumalacarregui:2013pma,Gleyzes:2014dya} and also studied in \cite{Gleyzes:2014qga,Lin:2014jga,Deffayet:2015qwa,Crisostomi:2016tcp}. Later, these theories were further understood  and generalized under the degeneracy criterion \cite{Langlois:2015cwa}.
Basically, a higher order scalar-tensor theory still propagates 3 degrees of freedom (DOF) if, in addition to the usual Hamiltonian and momentum 
constraints associated with diff invariance, it admits another primary constraint\footnote{Due to Lorentz invariance, this primary constraint usually leads to a secondary constraint~\cite{Langlois:2015skt, Crisostomi:2017aim}. However this is not the case for mimetic gravity \cite{Chamseddine:2013kea}, in which case the primary constraint is first-class and generates an extra symmetry.}.
These theories, denoted as Degenerate Higher Order Scalar-Tensor (DHOST) theories (or also Extended Scalar-Tensor (EST) theories), were introduced in \cite{Langlois:2015cwa} and further analysed in \cite{Langlois:2015skt, Crisostomi:2016czh, Achour:2016rkg, deRham:2016wji}.
A complete classification up to cubic order in second derivatives of the scalar field is given in \cite{BenAchour:2016fzp}. Their cosmological perturbations, in the framework of the Effective Theory of Dark Energy (see e.g. \cite{Gleyzes:2014rba}), are studied in \cite{Langlois:2017mxy}.
Analogously, similar constructions for vector interactions were introduced in \cite{Heisenberg:2016eld} and a classification for degenerate vector-tensor theories up to quadratic order was given in \cite{Kimura:2016rzw}.
 
The \Ost problem and the notion of degeneracy (necessary to avoid such a problem)
were systematically studied in the context of classical mechanics in \cite{Motohashi:2016ftl, Klein:2016aiq} and later in the context of higher order field theories without gauge symmetries in \cite{Crisostomi:2017aim}.
A similarly rigorous analysis however is still missing for field theories that possess gauge symmetries, such as gravity theories enjoying diff invariance. In this paper we attempt a first step in this direction.

\subsection{\Ost instabilities and constraints}
\label{crit}
\no Before presenting the content of our paper, let us briefly discuss our present understanding concerning the presence of \Ost modes in a  field theory. We follow the results of \cite{Motohashi:2016ftl, Klein:2016aiq, Crisostomi:2017aim} and underline some difficulties to extend them to diff invariant theories (see also \cite{Chen:2012au} as an alternative way to deal with \Ost modes).
In general there is a potential \Ost mode for each field in the action appearing with second time derivatives. 
In order to remove {\it all} of them, as a first requirement, we need a set of primary constraints equal in number to the fields that appear with second time derivatives.
In case we have fewer primary constraints, then \Ost modes, at least as many as the number of missing primary constraints, propagate in the theory.

Although these modes lead to instabilities in absence of extra symmetries, in the case of diff invariance for instance, they can be healthy.
A  well known example is $f(R)$ where the higher derivative mode described by the trace of the 3-dimensional metric is left unconstrained and leads to a propagating extra degree of freedom. 
In this case however this mode is perfectly healthy as can be seen by reformulating the theory as a standard scalar-tensor one with no higher-order derivatives at all.

Having the primary constraints however is not enough, {\it each} of them has to generate a secondary constraint, when evolved over time, in order to remove the \Ost mode associated (we do not discuss here the very special case where the primary constraints are first-class).
It is indeed upon exploiting the secondary constraint that the linear momentum in the Hamiltonian -- the characteristic signature of \Ost instability -- is removed.
When a primary constraint does not generate a secondary one, then the Hamiltonian is still left unbounded from below rendering the theory unstable.
Again, also this point could have loopholes when applied to gauge invariant theories, although we do not know any explicit counter-example showing its failure.

Therefore, bearing in mind all these subtleties that certainly deserve a deeper investigation, in this paper we retain a conservative approach and also consider  theories with fewer primary constraints (as in the case of $f(R)$) but, if there is not a secondary constraint generated by each primary one, then the theory is potentially unhealthy.

\subsection{From degenerate metric theories   to ``Chiral Scalar-Tensor theories''}

\no In this paper we begin exploring higher order, diff invariant, pure metric theories in a 
four dimensional space-time which are degenerate and discuss whether they appear to be free (or not) of \Ost modes.  We restrict ourselves to the case where the Lagrangian depends
at most on second derivatives of the metric.
In this context we  recover Chern-Simons gravity \cite{Jackiw:2003pm} as a partially degenerate theory\footnote{The definition
of partially degenerate higher order metric theories is given below equation \eqref{kinetic matrix}.} and analyse its number of degrees of freedom in full generality.
Inspired by this parity violating theory of gravity, we extend our analysis and construct new scalar-tensor theories with the same feature. We dub these theories ``Chiral Scalar-Tensor theories''.
Although they might be pathological  in their covariant form, the \Ost modes disappear in the unitary gauge (where the scalar field  depends on time only)  and the restricted version of these theories therefore makes sense as Lorentz breaking theories  similar to Horava-Lifshitz.

The paper is organised as follows. In Section \ref{DMT}, we study
four dimensional  diff invariant pure metric theories that are degenerate.
We start with fully degenerate Lagrangians and continue with a large class of  partially degenerate theories.
In Section \ref{CSTT}, we introduce the notion of  chiral scalar-tensor theories and find new classes of theories which violate parity and propagate only three degrees of freedom in the unitary gauge. We draw our conclusions in Section \ref{Conc}.

\section{Degenerate Metric theories}
\label{DMT}

\subsection{Action and ADM decomposition}
\label{ADMsec}

\no We consider the general action
\bea
S[g_{\mu\nu}] = \int d^4x \, \sqrt{-g} \, L \left(g_{\mu\nu}, \,\partial_\rho g_{\mu\nu}, \, \partial_{\rho} \partial_\sigma g_{\mu\nu} \right) \,,
\eea
governing the dynamics of the four-dimensional metric $g_{\mu\nu}$.
This action is assumed to depend at most on the second derivatives of the metric and,
due to Thomas' replacement theorem \cite{Thomas} (see also \cite{Horndeski:2017rtl} for a modern version),
the derivatives of the metric enter the Lagrangian through the Riemann tensor~$R_{\mu\nu\rho\sigma}$; this also guarantees  diffeomorphism (diff) invariance.
The action can thus be constructed by contracting the three following building blocks: the Riemann tensor, the metric and the Levi-Civita tensor\footnote{Note that the Levi-Civita tensor is defined by $\varepsilon^{\mu\nu\rho\sigma} =  \epsilon^{\mu\nu\rho\sigma}/{\sqrt{-g}} $ where $\epsilon^{\mu\nu\rho\sigma}$ is the fully antisymmetric symbol which takes value in $\{-1,0,+1\}$.} $\varepsilon^{\mu\nu\rho\sigma}$. 

In order to perform a Hamiltonian  analysis of the system, we need to separate space and time. We therefore foliate the space-time manifold $\cal M$ as $\Sigma \times \mathbb R$ 
and introduce the unit  time-like vector $n^\mu$ orthogonal to $\Sigma$, thus satisfying the normalization condition $n_\mu n^\mu=-1$.  This induces a three-dimensional metric on $\Sigma$ defined by $\g_{\mu\nu}\equiv g_{\mu\nu}+n_\mu n_\nu$.
Let us then consider the time direction vector $t^\mu \partial/\partial x^\mu \equiv \partial/\partial t$ (i.e. $t^\mu=(1,0,0,0)$) associated with a time coordinate $t$ that labels the slicing of spacelike hypersurfaces.
One can always decompose such a vector as $t^\mu =N n^\mu +N^\mu$,
thus defining the lapse function $N$ and the shift vector $N^\mu$ orthogonal to $n^\mu$. The time derivative (indicated with a dot) of spatial tensors is defined as the spatial projection of their Lie derivative with respect to $t^\mu$. In the following we will use latin indices ($i,j,k,\cdots$) to denote 3-dimensional objects living on the hypersurface $\g_{\mu\nu}$.
 
The ADM decomposition of the metric gives 
\bea\label{ADM}
g_{\mu\nu}=
\left(
\begin{array}{cc}
-N^2+\gamma_{ij}N^i N^j & \quad \gamma_{ij}N^j
\\
\gamma_{ij}N^i & \quad \gamma_{ij}
\end{array}
\right)\,,
\eea
and the components of the Riemann tensor in terms of the ADM variables are
(see for instance~\cite{Deruelle} where a Hamiltonian analysis of $f(\text{Riemann})$ was presented)
\bea\label{Rdecomp}
{\cal R}_{ij} \equiv n^\m n^\n R_{\m i \n j} &=& -\frac{1}{N} (\dot{K}_{ij} - {\cal L}_{\vec{N}} K_{ij}) + K_{i\l}K^\l{}_j + \frac{1}{N} D_i D_j N \, , \label{R1}\\
n^\m R_{\m ij\l} &=& D_\l K_{ij} - D_j K_{\l i} \, ,\label{R2}\\
R_{ij\l m} &=&  K_{i\l} K_{jm} - K_{im} K_{j\l} + {}^{(3)} R_{ij\l m} \, .\label{R3}
\eea
{The components on the LHS of the above equations are the bulk curvature components projected onto the surface $\Sigma$.}
We have used the notation ${}^{(3)} \!R_{ij\l m}$ for the three-dimensional Riemann tensor, ${\cal L}_{\vec{N}}$ for the Lie derivative along $N^i$,
$D_i$ for the covariant derivative compatible with $\gamma_{ij}$ and $K_{ij}$ for the components of the
 extrinsic curvature tensor defined by
\bea\label{extrinsic}
K_{ij} \; = \; \frac{1}{2N} \left( \dot{\gamma}_{ij} - D_i N_j - D_j N_i \right)\, .
\eea

Second time derivatives appear only for the spatial metric components $\g_{ij}$, and only in ${\cal R}_{ij}$ via the time derivative of the extrinsic curvature $\dot{K}_{ij}$. Notice that the same term is also the only one which  contains time derivatives of the lapse and shift.
Therefore, according to the \Ost  analysis, for a generic Lagrangian one could expect as many \Ost modes as the number of components of~$\gamma_{ij}$.

A necessary (but clearly not sufficient) condition to get rid of all of them, or part of them, is  that the theory has
 constraints in addition to the usual constraints associated with diff invariance.
It means that the Hessian matrix of the Lagrangian with respect to the second time derivatives of the spatial metric
\bea\label{kinetic matrix}
{\cal A}^{ij,\l m}(x,y) \; \equiv \;  \frac{\partial^2 L}{\partial \dot{K}_{ij}(x) \partial \dot{K}_{\l m}(y)}
\; = \; 4N(x)N(y)
\frac{\partial^2 L}{\partial \ddot{\gamma}_{ij} (x)\partial \ddot{\gamma}_{\l m}(y)} \,,
\eea
is degenerate\footnote{Note that the above matrix is local, i.e. ${\cal A}^{ij,\l m}(x,y) \propto \delta(x-y)$, because the Lagrangian contains at most second derivatives.}. According to the rank of the above matrix we will have a different number of primary constraints.
In this paper we study in detail only two cases: Lagrangians associated with a Hessian matrix of rank 0 (fully degenerate case) and of rank 1 (partially degenerate case). We will also briefly 
discuss in Appendix \ref{higherank} the case of larger ranks, leaving the detailed analysis for future works.
If, by contrast, the Hessian matrix $\cal A$ is invertible, then the theory propagates 8 degrees of freedom, some 
of them being necessarily ghosts (see \cite{Bonifacio:2015rea} for the linear analysis).

\subsection{Fully degenerate theories}

\no In this section we study all the theories that satisfy ${\cal A}^{ij,\l m} = 0$, which implies that their
Lagrangian is linear in $\ddot{\gamma}_{ij}$.
Requiring the Lagrangian to be linear in second time derivatives, means that the corresponding equations of motion  can be at most of third order.

\subsubsection{Degenerate Lagrangians}
\no In a pioneering paper \cite{Lovelock}, Lovelock already classified all the possible Lagrangians satisfying this condition and showed that there are only 3 independent terms in addition to the usual Einstein-Hilbert Lagrangian $R$:
\bea\label{fully}
GB \equiv (\star R^{\mu\nu}{}_{\alpha\beta})(\star R^{\alpha\beta}{}_{\mu\nu}) \, , \quad
P \equiv (\star R^{\mu \nu}{}_{\alpha\beta}) R^{\alpha\beta}{}_{\mu\nu} \, , \quad
C \equiv  ( \star R^{\mu\nu}{}_{\rho\sigma} ) ( \star R^{\rho\sigma}{}_{\alpha\beta} )( \star R^{\alpha\beta}{}_{\mu\nu} ) \,, 
\eea
where $\star$ holds for the Hodge dual
\bea
\star R^{\mu\nu}{}_{\rho\sigma} \; \equiv \; \varepsilon^{\mu\nu\alpha\beta} R_{\alpha\beta\rho\sigma} \, .
\eea
The Ricci scalar (R) gives second order field equations; the  Gauss-Bonnet (GB) and the Pontryagin (P) terms are topological invariants (in 4 dimensions) and 
their variation yields no term to the field equations~\cite{Lanczos:1938sf}. The last curvature invariant (C) is the only one whose equations of motion are of third order\footnote{We thank Alex Vikman for bringing to our attention this unique cubic term as well as the article \cite{Lovelock}.}.
 
Linearity in $\ddot{\gamma}_{ij}$ translates into linearity in ${\cal R}_{ij}$, given in (\ref{R1}), and from this respect it is easy to  understand Lovelock's result.
Indeed the Ricci scalar is the only density which is linear in the Riemann tensor while at the quadratic and cubic levels we need to make use of the $\varepsilon$ tensor to avoid non-linearities in ${\cal R}_{ij}$, which leads uniquely to $GB$, $P$ and $C$. Finally, even using the $\varepsilon$ tensor, it is not possible to avoid at least quadratic terms in ${\cal R}_{ij}$ when one considers more than 3 powers of the Riemann tensor. 

Therefore, in addition to GR, there is only one other non trivial fully degenerate Lagrangian, namely $C$.  
Since this term leads to third order EOM it has never attracted much attention in the literature and a canonical analysis to count its number of DOF, as well as their stability, is still missing, to the best of our knowledge. In the rest of this section we partially fill this gap and perform the Hamiltonian  analysis of this theory.

\subsubsection{Hamiltonian analysis of $C$}
\label{Csec}

\no Using equations \eqref{R1}, \eqref{R2} and \eqref{R3},  the  action
\beq
S_C =  \int d^4x \sqrt{-g} \, C \,, \label{cubic action}
\eeq
can be rewritten in the form
\bea\label{decomposition}
S_C = \int d^4x \, \left( \dot{K}_{ij} \Pi^{ij} -  V  \right) \,,
\eea
where the 3-dimensional rank 2  density $\Pi^{ij}$  is defined by
\bea
\Pi^{ij} \; \equiv \; \sqrt{-g} \, \frac{\partial C}{\partial \dot K_{ij}} \, ,
\eea
and the ``potential" $V$ contains all the other terms from the decomposition of (\ref{cubic action}) that do not involve $ \dot{K}_{ij}$.
The explicit form of $V$ is quite long and we do not reproduce it here, as  only some of its general properties will be useful in the following.
An important property is that, after several integrations by parts, we can rewrite the potential $V$ as
\beq
V = N \, V_0 \, + \, N^i V_i \,,
\eeq
where $V_0$ and  $V_i$, like $\Pi^{ij}$, depend only on $\gamma_{ij}$, $ {}^{(3)} R_{ij\l m}$, $K_{ij}$ and their spatial derivatives and do not depend explicitly on the lapse function and shift vector, which enter only through the extrinsic curvature. This can be seen  as a consequence of the diff invariance of the action (\ref{cubic action}). 

Since the action involves second time derivatives of the spatial metric $\g_{ij}$ in $\dot{K}_{ij}$, it is convenient to consider the following equivalent form
\bea\label{equivQ}
S_{eq}  =  \int d^4x \,  \left[ \dot{Q}_{ij} \Pi^{ij} - N \, V_0 \, - \, N^i V_i 
  +  2  N p^{ij} \left( K_{ij} - Q_{ij}\right) \right] \,,
\eea
where we have  introduced the new 3-dimensional symmetric tensors $Q_{ij}$ and $p^{ij}$ in order to make the Lagrangian depend explicitly on first time derivatives only. 
The equations of motion for $p^{ij}$ enforce the condition $Q_{ij}=K_{ij}$, recovering therefore the original action (\ref{decomposition}).
Note that in \eqref{equivQ}, $\Pi^{ij}$, $V_0$ and $V_i$  now depend on $Q_{ij}$ and not  on $K_{ij}$.
In this form,  the action has a  linear dependence on the  lapse and the shift, which  clearly appear as  Lagrange multipliers in this reformulation.
Indeed, expanding the last term in \eqref{equivQ} and integrating by parts, we get
\bea\label{first order form}
S_{eq}  =  \int d^4x \,  \left( \dot{Q}_{ij} \Pi^{ij} + \dot{\gamma}_{ij} p^{ij} - N {\cal H}_0 - N^i {\cal H}_i \right) \,,
\eea
where 
\bea
{\cal H}_0 \; \equiv \; V_0  +2 p^{ij}Q_{ij} \,, \qquad \text{and} \qquad
{\cal H}_i \; \equiv \; V_i - 2 D^j p_{ij} \, . \label{H0Hi}
\eea

We are now ready to perform the Hamiltonian analysis starting in a phase space endowed  with the following 16 pairs of conjugate variables
\bea
&& \{\gamma_{ij}(x), p^{kl}(y)\} = \{Q_{ij}(x), P^{kl}(y) \} = \frac{1}{2 }(\delta_{i}^k \delta^l_{j} + \delta_{j}^k \delta^l_{i} ) \, \delta^{(3)}(x-y) \, ,\label{phase space}\\
 &&  \{N(x), \pi^0(y) \} = \, \delta^{(3)}(x-y) \, ,\qquad  \{N^i(x), \pi_j(y) \} =\delta^i_j \, \delta^{(3)}(x-y) \; , \nonumber
\eea
where $\delta^{(3)}(x-y)$ denotes the  Dirac delta distribution on the space hypersurface. 
The fact that $\gamma_{ij}$ and $ p^{ij}$ are conjugate variables is 
manifest 
from \eqref{first order form}. 

\medskip

As the action does not involve time derivatives of the lapse and the shift, we recover  the usual four primary constraints
\bea\label{primary1}
\pi_\mu \; \approx \; 0 \qquad (\mu = 0,1,2,3) \, ,
\eea
which, in analogy with the Hamiltonian formulation of GR, are closely related to the diffeomorphism invariance  of the theory. Furthermore, since the Lagrangian is fully degenerate, computing the conjugate momenta $P^{ij}$ leads to 6 additional primary constraints
\bea\label{primary2}
\chi^{ij} \; \equiv \; P^{ij}  - \Pi^{ij} \; \approx \; 0 \, .
\eea
Hence, the total Hamiltonian of the theory takes the form
\bea
H_{T} = H_C + \int d^3 x  \, \left( \xi_\mu \pi^\mu + \xi_{ij} \chi^{ij}  \right) \,, \qquad 
H_C \equiv \int d^3 x \,  \left( N {\cal H}_0 + N^i {\cal H}_i \right) \,,
\eea
where $\xi_\mu$ and $\xi_{ij}$ are Lagrange multipliers that enforce the primary constraints \eqref{primary1} and \eqref{primary2}.

Requiring the time conservation
of \eqref{primary1} leads to the following secondary constraints
\bea\label{first class}
{\cal H}_0 \, \approx \, 0 \, , \qquad {\cal H}_i \, \approx \, 0 \, .
\eea
These constraints are closely related to the usual Hamiltonian and momentum constraints, which generate space-time diffeomorphisms. More precisely, they are first class up to the addition of the other second class constraints.

On the other hand, requiring the conservation in time of \eqref{primary2} leads to the equation
\bea
\{\chi^{ij}(x) ,\, H_C \} + \int d^3 y  \, \{\chi^{ij}(x), \chi^{k\l}(y)\} \xi_{k\l}(y) \approx 0 \, . \label{nosec}
\eea
Furthermore, the Dirac matrix between the constraints $\chi^{ij}$, defined by
\bea
\Delta^{ij,k\l}(x,y) \; \equiv \;  \{\chi^{ij}(x), \chi^{k\l}(y)\} = \frac{\de \Pi^{k\l}(y)}{\de Q^{ij}(x)} - \frac{\de \Pi^{ij}(x)}{\de Q^{k\l}(y)}  \,,
\label{DeltaC}
\eea
turns out to be invertible and therefore we can use equation (\ref{nosec}) to
fix the Lagrange multipliers $\xi_{ij}$ in terms of the phase space variables. 
As a consequence, there are no secondary constraints associated with (\ref{primary2}). Hence the Dirac analysis closes with the 8 first class constraints \eqref{primary1} and 
\eqref{first class} together with the  6 second class constraints \eqref{primary2}. This results in $[32-(8 \times 2) - 6]/2 = 5$ degrees of freedom.
Note that adding the Einstein Hilbert action to \eqref{cubic action} does not change the conclusion of the Hamiltonian analysis: we end up
with 5 degrees of freedom in total.

To conclude, let us notice that ${\cal H}_0$ and ${\cal H}_i$ in (\ref{H0Hi}) are linear in $p^{ij}$ and therefore the Hamiltonian appears unbounded from below. This is the  characteristic feature of \Ost instabilities, indicating that the extra 3 DOF are likely to be ghosts. These extra  DOF could be eliminated if  secondary constraints were present,  thereby removing the linear dependence of the Hamiltonian  on the momenta associated with the higher derivative modes. In the present case the absence of secondary constraints  suggests that the extra modes are not stable.
In Appendix \ref{Cpert}, we confirm the instability of \eqref{cubic action} at  linear order in perturbation theory.

\subsection{Partially degenerate theories}
\label{pdt}
\no In this section, we study  theories with  a Hessian matrix (\ref{kinetic matrix}) of rank 1.
A straightforward method to construct models of this type simply consists in considering generic functions of the fully degenerate Lagrangians\footnote{One can suspect that  all partially degenerate Lagrangians for a metric with a rank 1 Hessian matrix are
of the form  \eqref{f(Y)}, although we have no formal proof for this.}  studied in the former section
\bea
S \; = \; \int d^4 x \sqrt{-g} \, f(Y) \,, \qquad Y = R, GB, P \,. \label{f(Y)}
\eea
Indeed, the linearity argument concerning $\ddot \g_{ij}$ ensures that, when $f''\neq 0$, the kernel of $\cal A$ is of co-dimension 1, which means that the theory admits $(6-1)$ primary constraints.
Since we have already discussed  the potential problems of the C term, we will not consider $f(C)$ theories here and we will concentrate our attention on  $f(GB)$ and $f(P)$, the case of $f(R)$ being already well known.
In Appendix \ref{higherank} we also give a  short discussion about theories with a Hessian matrix~(\ref{kinetic matrix}) of rank higher than 1.

\subsubsection{General 
discussion}
\no The Lagrangians $f(R)$ and $f(GB)$ are well known to define theories that propagate 3 DOF and are equivalent{\footnote{By equivalent we mean that they have identical classical equations of motion in vacuum.}} to scalar-tensor theories within the class of Horndeski. This fact has been known for a long time for $f(R)$ (see e.g. \cite{Sotiriou:2008rp, DeFelice:2010aj} for reviews on $f(R)$ theories): the action can be rewritten as a  Brans-Dicke-like theory. The equivalence of $f(GB)$ with Horndeski is more recent and was established  only at the level of the equations of motion\footnote{However, the Horndeski form of $f(GB)$  involves a logarithmic function of $X$ (where $X$ is the kinetic term for the scalar field, $X\equiv \de_\m \phi \de^\m \phi$), signalling a non-analyticity issue when $X \rightarrow 0$, something that is not very evident in $f(GB)$. We will clarify the origin of this singular point in the next subsection.} \cite{Kobayashi:2011nu}. 

Finally, the last theory, $f(P)$, 
can be related to Chern-Simons gravity \cite{Jackiw:2003pm}, which has been much  studied in the literature  (see \cite{Alexander:2009tp} for a review). Indeed, repeating the same procedure that transforms $f(R)$ into a scalar-tensor theory (see for example \cite{Sotiriou:2008rp, DeFelice:2010aj}), action (\ref{f(Y)}) can be rewritten as
\beq
S  =  \int d^4 x \sqrt{-g} \left[ \phi \, Y - U(\phi) \right] \,, \label{f(Y)equiv}
\eeq
where $U(\phi)$ is a potential given by
\beq
U(\phi) = \psi(\phi) \phi - f(\psi(\phi)) \,, \qquad \phi \equiv f_{\psi} (\psi) \,.
\eeq
The reformulation (\ref{f(Y)equiv}) will be useful for our analysis of the various cases considered below.

\subsubsection{$f(GB)$ theory}
\label{fGB}
\no In order to exploit  
our previous analysis of  fully degenerate theories, it is convenient to
 study $f(GB)$ in the equivalent form  (\ref{f(Y)equiv}), i.e. 
\beq
S  =  \int d^4 x \sqrt{-g} \, \left[\phi \, GB \, - \, U(\phi)\right] \, ,\label{f(GB)}
\eeq
where the potential $U$ will be ignored in the following, as its presence  does not modify the conclusion.

Using the ADM decomposition of section \ref{ADMsec}, we can apply the same strategy used for studying the $C$ term in section \ref{Csec}.
All we need to do is 
to construct, for  the Gauss-Bonnet action, the analogs of  $\Pi^{ij}$ and $V$ defined previously, 
and introduce an extra pair of conjugate variables to account for the scalar field $\phi$, i.e. 
\bea
\{\phi(x), \pi_\phi(y) \} \, = \, \delta^{(3)}(x-y)\, .
\eea
For the action (\ref{f(GB)}) the total Hamiltonian takes the form
\bea
H_{T} = H_C + \int d^3 x  \, \left( \xi_\mu \pi^\mu + \xi_{ij} \chi^{ij}  + \lambda \, \pi_\phi \right) \,, \qquad 
H_C \equiv \int d^3 x   \, \left( N {\cal H}_0 + N^i {\cal H}_i \right) \,, \label{HTfGB}
\eea
where 
\bea
{\cal H}_0 \; \equiv \; V_0  + 2 \, p^{ij}Q_{ij} \,, \qquad \text{and} \qquad
{\cal H}_i \; \equiv \; V_i - 2 \, D^j p_{ij}  \, , \label{HfGB}
\eea
and now $V_0, V_i$ involve also the scalar field $\phi$ and its space derivatives.  We avoid to report their explicit form here, they are however straightforward to compute.
We have also introduced in the total Hamiltonian the Lagrange multiplier $\lambda$ to enforce the new primary constraint $\pi_\phi \approx 0$, since the action (\ref{f(GB)}) does not contain any kinetic term for the scalar field.
Notice that  it is the linearity in  $p^{ij}$ of the Hamiltonian (\ref{HfGB}) which is potentially dangerous. 

Concerning the study of the stability under time evolution of the primary constraints $\pi^\mu \approx 0$, the same arguments of section \ref{Csec} apply, and they generate the secondary constraints ${\cal H}_\mu \approx 0$. They are first class (up to adding second class constraints).

The degeneracy of the Hessian matrix leads to 6 primary constraints $\chi^{ij}$ given by
\bea\label{primary2GB}
\chi^{ij} \; \equiv \; P^{ij}  - \phi \, \Pi^{ij} \; \approx \; 0 \, ,
\eea
where $\Pi^{ij}$ are obtained from the GB term 
\bea
\Pi^{ij} = 4  \sqrt{\gamma} \, \left[  2 \left( Q^{ik} Q_k^j - {}^{(3)}R^{ij} - Q Q^{ij} \right) + \g^{ij} \left( {}^{(3)}R + Q^2 - Q_{k \l} Q^{k \l} \right) \right] \,. \label{PiGB}
\eea
Making use of the explicit form of $\Pi^{ij}$ in (\ref{PiGB}), it is easy to compute the Dirac matrix between the constraints $\chi^{ij}$ and show that it identically vanishes
\bea
\{\chi^{ij}(x), \chi^{k\l}(y)\} = \phi(y) \frac{\de \Pi^{k\l}(y)}{\de Q^{ij}(x)} - \phi(x) \frac{\de \Pi^{ij}(x)}{\de Q^{k\l}(y)} = 0 \,.
\eea

Let us now see whether secondary constraints arise.
First, requiring the stability under time evolution of \eqref{primary2GB} leads  to
\bea
\{\chi^{ij} ,\, H_C \} - \lambda \,  \Pi^{ij}  \approx 0 \, . \label{5secGB}
\eea
Taking the trace of (\ref{5secGB}) enables one to determine the Lagrange multiplier $\lambda$ in terms of the phase space variables. 
The traceless part gives 5 secondary constraints. We then consider the constraint $\pi_\phi \approx 0$ whose time evolution yields
\beq
\{ \pi_\phi  , H_C \} +  \Pi^{ij} \xi_{ij} \approx 0 \,,
\eeq
which can be  solved to write  the trace of $\xi_{ij}$ in terms of the canonical variables and the remaining 5 components of the traceless part
of $\xi_{ij}$.

Finally, the evolution of the 5 secondary constraints given by the traceless part of (\ref{5secGB}) determines the traceless component of $\xi_{ij}$ and the analysis stops.

Therefore, starting with 32 (metric) + 2 (scalar) canonical variables, and having 8 first class  and 12 (7 primary + 5 secondary) second class constraints, we end up with a total of 3 DOF, which is compatible with the equivalence of $f(GB)$ with a scalar-tensor theory. However, here, the 6 primary constraints (\ref{primary2GB}) coming from the higher derivative modes in the action, generate only 5 secondary constraints and the Hamiltonian still remains linear in the trace of the momentum $p^{ij}$. This seems to indicate that the theory possesses one \Ost mode. Note that this \Ost mode could be removed by   adding to the action (\ref{f(GB)}) a kinetic term for the scalar field so that  the primary constraint $\pi_\phi \approx 0$ disappears from the total Hamiltonian and the 6 primary constraints (\ref{primary2GB}) generate  6 secondary constraints. This does not change the total number of DOF, but makes the theory \Ost free by removing any linear  momentum dependence. This suggests that only $f(GB)$ supplemented with an explicit kinetic term for the scalar field is classically equivalent  to some Horndeski theory\footnote{This may explain the origin of the singular $X \rightarrow 0$ limit of the Horndeski formulation of $f(GB)$ (shortly reported in footnote 8): the theory does not allow a vanishing kinetic term for the scalar field.}.
\medskip

Notice that the  same argument a priori seems to apply to $f(R)$ too, suggesting the (erroneous) conclusion that  $f(R)$ needs an explicit kinetic term for the scalar field in order to be ghost free.
This is obviously not the case and we believe the reason lies in the very special structure of this theory.
Indeed a conformal transformation, performed on the equivalent formulation (\ref{f(Y)equiv}) of the theory, removes the coupling between the metric and the scalar field which acquires its own kinetic term. A similar transformation does not seem to exist for $f(GB)$.

\subsubsection{$f(P)$ -- Chern-Simons gravity}

\no 
The reformulation (\ref{f(Y)equiv}) shows that  $f(P)$ is related, up to a potential term, to non-dynamical Chern-Simons, whose action reads
\beq
S_{CS}  =  \int d^4 x \sqrt{-g} \, \phi \, P \,. \label{ndCS}
\eeq
Chern-Simons modification of gravity is usually seen as 
an effective field theory (EFT), truncated at quadratic order in the curvature, in a low-energy expansion of a more fundamental theory~\cite{Alexander:2009tp}.
Indeed, since it leads to equations of motion with higher-order derivatives, 
it is expected to contain \Ost modes if treated as a complete theory (i.e. not as a perturbative expansion).
For the so-called dynamical Chern-Simons gravity (where also an explicit kinetic term for $\phi$ is present), \cite{Dyda:2012rj} showed that there is at least a ghost instability above a certain momentum cutoff and \cite{Delsate:2014hba} provided evidence that the theory does not admit a well-posed initial value formulation (see also \cite{Okounkova:2017yby} for numerical simulations using the perturbative approach).
However, to the best of our knowledge, a proper canonical analysis of this theory has never been performed in order to count  the number of DOF at the  nonlinear level. In the following, we present a canonical analysis of non-dynamical Chern-Simons gravity (\ref{ndCS}), then we add a potential $U$ to study $f(P)$. Finally, we also add explicitly a kinetic term for $\phi$ in order to analyse the dynamical Chern-Simons gravity. 
\\

\paragraph{Non-Dynamical Chern-Simons gravity \\}
 
\no \\Using the decomposition of the Riemann tensor given in \eqref{R1}, \eqref{R2} and \eqref{R3}, and the equivalent first-order formulation of the action, the Pontryagin tensor gives
\beq
\Pi^{ij} = 8 \left( \e^{ik\l}D_\l Q^j_k + \e^{jk\l}D_\l Q^i_k \right) \,, \label{PiP}
\eeq
and
\bea
V &=& 8 \, \e^{ijk} \left[  2 \left( {\cal L}_{\vec{N}} Q_{i\l} + D_i D_\l N  \right) D_k Q^\l _j \right.\nb \\
&+& \left. N \left(   2\,  Q_{i \l} Q^{\l m} D_k Q_{jm} -
2\, Q_{i \l } Q^m_j D_m Q^\l _k - 
 {}^{(3)}R_{jk \l }{}^m D_m Q^\l _i
\right)
\right] \,, \label{VCS}
\eea 
where we have used $\varepsilon^{ijk}=\epsilon^{ijk}/\sqrt{\gamma}$.

The above $\Pi^{ij}$ and $V$ satisfy two important properties, related to the invariance of the action (\ref{ndCS}) under conformal transformations.
First $\Pi^{ij}$ is traceless, meaning that the action does not contain time derivatives of the trace of $Q_{ij}$, $Q\equiv\gamma^{ij}Q_{ij}$.
Second, one can check that the dependence of the potential $V$ on $Q$ is at most linear, meaning that $Q$  effectively plays the role of  a Lagrange multiplier, similarly to the lapse and shift.
It is  therefore useful to explicitly decompose any tensor into its trace and traceless components: we drop the indices to indicate the trace and use a tilde to denote the traceless part.

The total Hamiltonian takes the form
\bea
&& H_{T} = H_C + \int d^3 x \left( \xi_\mu \pi^\mu + \xi \, P + \tilde \xi_{ij} \tilde \chi^{ij}  + \lambda \, \pi_\phi \right) \,, \label{HTCS} \\
&& H_C \equiv \int d^3 x \left( N {\cal H}_0 + N^i {\cal H}_i + N Q {\cal H}_c \right) \,, 
\eea
where 
\bea
&& {\cal H}_0 \; \equiv \; V_0  + 2 \, \tilde p^{ij} \tilde Q_{ij} \,, \qquad 
{\cal H}_i \; \equiv \; V_i - 2 \left( D_j  \tilde p_i^j  + \frac{D_i p}{3} \right)\, , \\[2ex]
&& {\cal H}_c \equiv 2  \left( 8 \, \phi \, \e^{jk\l} \tilde Q_{ij} D_\l \tilde Q^i_k + \frac{p}{3} \right) \,. \label{secFC}
\eea
The 6 primary constraints $\chi^{ij} \approx 0$  (\ref{primary2GB}) can be  divided into trace and traceless parts:
\bea\label{primaryP}
P \approx 0 \,, \qquad \tilde \chi^{ij} \; \equiv \; \tilde P^{ij}  - \phi \, \tilde \Pi^{ij} \; \approx \; 0 \, .
\eea

The time evolution of the primary constraint $P \approx 0$ leads to the secondary constraint
\beq
{\cal H}_c  \approx 0 \, \label{secFC}\, .
\eeq
The evolution of the constraints $\pi^\mu \approx 0$, yields, as usual, the secondary constraints 
\bea
{\cal H}_0 \approx 0 \, , \quad {\cal H}_i \approx 0\,.
\eea
Before investigating the time evolution of $\tilde \chi^{ij}\approx 0$, it is useful to  compute the Dirac matrix associated with these constraints, which is given by
\bea
\tilde \Delta^{ij,k\l}(x,y)\equiv \{\tilde \chi^{ij}(x), \tilde \chi^{k\l}(y)\} = 4 ( \e^{ikm} \, \g^{j\l} + \e^{jkm} \, \g^{i\l} + \e^{i\l m} \, \g^{jk} + \e^{j\l m} \, \g^{ik})   \phi_m \, \d^{(3)} (x-y) \,, \nb \\ \label{DMCS}
\eea
where $\phi_m \equiv D_m \phi$. Assuming that $\phi_i$ is non zero, one sees that the symmetric matrix $(\phi_k \phi_\l)$ is a null eigenvector of the Dirac matrix, i.e.
\bea
\tilde \Delta^{ij,k\l} \phi_k \phi_\l =0 \, .
\eea
At this stage, it is useful to introduce the projector orthogonal to $\phi_i$
 \bea
\hat \gamma^i_j \equiv \gamma^i_j - \frac{\phi^i \phi_j}{\phi_k\phi^k} \,;
\eea
the projection orthogonal to $\phi_i$ of any 3-dimensional tensor will be denoted with a hat in the following.

Let us now return to  the constraint analysis. Evolving the 5 primary constraints $\tilde \chi^{ij}$ and taking the projection  along the direction $(\phi_i\phi_j)$, one gets
\beq
\{\tilde \chi^{ij}, H_C \} \phi_i \phi_j - \lambda\,  \tilde \Pi^{ij} \phi_i \phi_j \approx 0 \,, \label{lambdaconst}
\eeq
which can be solved in general to determine  the Lagrange multiplier $\lambda$.
The projection along $\hat \g_{ij}$  determines the 4  Lagrange multipliers $\hat \xi_{ij}$ in terms of the canonical variables as the matrix $\hat \Delta^{ij,k\l}$ is invertible.
Finally, the time evolution of the constraint $\pi_\phi \approx 0$ yields the component $\tilde \xi_{ij} \phi^i \phi^j$ of the Lagrange multipliers $\tilde \xi_{ij}$ as
\beq
\tilde \xi_{ij} \phi^i \phi^j = - \frac{\{ \pi_\phi  , H_C \} + \hat \Pi^{ij} \hat \xi_{ij}}{\tilde \Pi^{ij} \phi_i \phi_j} \,.
\eeq

At this point we are left with the secondary constraints ${\cal H}_0, \, {\cal H}_i$ and ${\cal H}_c$ and the Lagrange multipliers $\xi_\mu$ and $\xi$ are still undetermined.
It is easy to see that the primary constraints $\pi^\mu\approx 0$ and $P\approx 0$ have vanishing Poisson brackets with all the other constraints, i.e. they are first class. By contrast,  it is a non-trivial task to show that  their associated secondary constraints ${\cal H}_0, \, {\cal H}_i$ and ${\cal H}_c$ are also first class (up to the addition of second class constraints)  and that the algebra closes. 
However, it is natural to expect that this is indeed the case since these constraints are associated with symmetries, namely the diffeomorphism and the conformal invariance of  the action (\ref{ndCS}), and we will assume so in the following\footnote{Notice that a similar result has been shown for the Weyl squared term in \cite{Boulware}, where ${\cal H}_c$ has been proven to be the generator of conformal transformations under which the theory is invariant.}.
In summary, we thus have  32 (metric) + 2 (scalar) canonical variables constrained by  10 first class and  6 second class constraints, leading to $[34 - (10 \times 2) - 6]/2 = 4$ degrees of freedom.

\medskip

This counting applies only to the action \eqref{ndCS}. If one adds a potential $U(\phi)$, as is necessary for  $f(P)$, or  the standard Einstein-Hilbert (EH) term, as in the case of Chern-Simons {\it modified gravity}, the total Lagrangian is no longer  conformally invariant. 
As a consequence,   the constraints $P$ and ${\cal H}_c$ become second class. This gives  one extra DOF in comparison with the above analysis, leading to a total of 5 degrees of freedom for $f(P)$ or for Chern-Simons modified gravity.

\medskip

As in the case of the fully degenerate theory (\ref{cubic action}), the primary constraints $\tilde \chi^{ij}$, associated with the higher derivative modes in the Lagrangian, do not generate secondary constraints, leaving therefore the  Hamiltonian linear in the momenta $\tilde p^{ij}$. According to our  discussion in the introduction, this potentially signals that the extra 2 or 3 DOF (depending on whether there is a conformal invariance or not) are \Ost modes and the theory is likely to be unstable. Note however that these modes could be ignored if one considers the Chern-Simons term as a perturbative correction to General Relativity in the EFT spirit.

\medskip

The absence of secondary constraints generated by $\tilde \chi^{ij}$ comes from the non vanishing of the Dirac matrix (\ref{DMCS}), due  to the presence of the spatial derivatives of $Q_{ij}$ in (\ref{PiP}).
In the so called unitary gauge, i.e. where the scalar field is by construction uniform, the Dirac matrix (\ref{DMCS}) vanishes and the evolution of the 5 primary constraints $\tilde \chi^{ij}$ leads to 5 extra secondary constraints, removing all the \Ost modes.
Considering therefore the unitary gauge expression of CS as a Lorentz breaking (different) theory, saves the day and represent a healthy parity violating extension of Horava--Lifshitz involving also $\dot K_{ij}$.
\\

\paragraph{Dynamical Chern-Simons gravity \\}

\no \\ To conclude our analysis of CS gravity, let us briefly discuss the case of dynamical CS gravity, defined by the action (\ref{ndCS}) supplemented with a kinetic term for the scalar field $\phi$ of  k-essence form for instance
\bea
\label{kinphi}
S_{\phi, \rm kin} \; \equiv \; \int d^4x \, \sqrt{-g} \, F(\phi,X) \,,
\eea
where $F$ is an arbitrary function with a non-trivial dependency on $X$ (i.e. $F_X\neq 0$).
In that case, the primary constraint $ \pi_\phi \approx 0$ disappears from the  Hamiltonian analysis, as a consequence
 we set $\lambda=0$ in the total Hamiltonian \eqref{HTCS} and equation (\ref{lambdaconst}) now becomes a secondary constraint
\beq
\{\tilde \chi^{ij}, H_C \} \phi_i \phi_j \approx 0 \,. \label{newsec}
\eeq
Remarkably, the Poisson brackets of this new constraint (\ref{newsec}) with $P$ and ${\cal H}_c$ do not vanish in general, 
making these latter second class and not anymore first class constraints.
This is not surprising since the kinetic term of $\phi$ breaks in general the invariance under conformal transformations of the original action 
\eqref{ndCS}.
From the evolution of ${\cal H}_c$ and the evolution of (\ref{newsec})
it is now possible to fix the component $\tilde \xi_{ij} \phi^i \phi^j$ of the multipliers $\tilde \xi_{ij}$ and the last multiplier $\xi$ that remained undetermined in the non-dynamical case.

The analysis therefore ends up  with 6 (primary) + 2 (secondary) second class constraints, in addition to the 8 first class constraints due to the diff invariance, resulting in a total of $[34 - 8 \times 2 - 8]/2 = 5$ DOF.
We thus obtain 
in the dynamical case as many  DOF as in non-dynamical Chern-Simons gravity plus the EH term\footnote{If the EH term is not included, in the very special case of a conformally invariant kinetic term 
$F(\phi,X)=f(\phi) X^2$  in \eqref{kinphi}, we still lose the primary constraint $ \pi_\phi \approx 0$, but the constraints $P\approx 0$ and ${\cal H}_c\approx 0$ remain first class due to the preserved conformal invariance of the action. In addition we have the 5 primary constraints $\tilde \chi^{ij}$ and the secondary constraint (\ref{newsec}) that are second class.
As a consequence, the total number of DOF is $[34 - (10 \times 2) - 6]/2 = 4$.}.
In the present case,  some of the primary constraints associated with the higher derivative modes in the Lagrangian do not lead to secondary constraints. This implies that the Hamiltonian is left linear in the components $\hat p^{ij}$ of the momentum $p^{ij}$, making the theory probably unstable.

\smallskip

\section{Chiral Scalar-Tensor Theories}
\label{CSTT}

\no Inspired by the analysis of $f(P)$ and Chern-Simons  gravity, in this last section we entertain the possibility of constructing healthy scalar-tensor theories, i.e. without \Ost modes, featuring parity violating effects.
For this purpose it is essential that the action involves an odd number of Levi-Civita tensors $\varepsilon^{\mu\nu\rho\sigma}$ and, for simplicity, 
we will restrict our attention to the cases where there is only one.

CS action (\ref{ndCS}) is the simplest scalar-tensor theory of this kind one can write down, but, given the structure of constraints revealed in the previous section, it is potentially unstable.
It is possible however to generalise this action by including first and second derivatives of the scalar field: $ \phi_\mu \equiv  \de_\m \phi$ and $ \phi_{\mu\nu}  \equiv \nabla_\mu \phi_\nu  $.
We will explore two types of extensions. In the first case, we consider Lagrangians involving only first order derivatives of $\phi$, which implies that the Lagrangians must be at least quadratic in the Riemann tensor. In the second case, we consider terms that are linear in the Riemann tensor while linear or quadratic in second derivatives of $\phi$.

\subsection{First derivatives of the scalar field only}

\no With only first derivatives of the scalar field, one cannot construct a Lagrangian that depends linearly on the Riemann tensor (and on the Levi-Civita tensor). With two Riemann tensors, one finds 
 four independent terms of this type:
\bea
&& L_1 \equiv \varepsilon^{\mu\nu\a\b} R_{\a\b\r\s} R_{\m\n}{}^\r{}_\lambda \phi^\s \phi^\lambda \,, \qquad
L_2 \equiv \varepsilon^{\mu\nu\a\b} R_{\a\b\r\s} R_{\m\lambda}{}^{\r\s} \phi_\n \phi^\lambda \,,  \nb \\
&& L_3 \equiv \varepsilon^{\mu\nu\a\b} R_{\a\b\r\s} R^\s{}_\n \phi^\r \phi_\m \,, \qquad\quad
L_4 \equiv X\, P \,,
\eea
where we recall that $X\equiv \phi_\mu \phi^\mu$ and $P \equiv \varepsilon^{\mu\nu\r\s} R_{\r\s\alpha\beta} R^{\alpha\beta}{}_{\mu\nu} $ is the Pontryagin term. In the following, we will analyse the linear combination
\beq
S = \sum_{A = 1}^4 \int d^4 x \sqrt{-g} \, a_A L_A\,, \label{chiral1}
\eeq
where $a_A(\phi,X)$ are a priori arbitrary functions of $\phi$ and $X$. 

\subsubsection{Brief Hamiltonian analysis}

\no To perform the Hamiltonian analysis of the action (\ref{chiral1}) we can rely on the same tools used in the previous sections, i.e. the ADM decomposition of equations \eqref{R1}, \eqref{R2} and \eqref{R3} together with the first order reformulation of the action. The only new ingredient we need is the decomposition of $\phi_\mu$, namely
\beq
\phi_\mu = \frac{1}{N} \left({\cal L}_{\vec{N}} \phi - \dot \phi \right)n_\m + D_\m \phi \,.
\eeq

One must now take into account two velocity terms, i.e. $\dot Q_{ij}$ and $\dot \phi$. Whereas the presence of the $\varepsilon$ tensor prevents terms quadratic in $\dot Q_{ij}$, it allows mixed terms in $\dot Q_{ij}$ and $\dot \phi$.
Therefore, in order to have the 6 primary constraints of the form (\ref{primary2}) that are the first necessary (but not sufficient) condition to remove the \Ost modes, the functions $a_A$ need to be tuned to avoid this coupling. 
This requirement leads to the conditions that the $a_A$ depend on $\phi$ only, and
\beq
a_1 = a_3 = 0 \qquad \text{and} \qquad a_2+4 a_4 = 0 \,. \label{cond1}
\eeq
One is left with only one free function, say $a_2\equiv f(\phi)$, and the action \eqref{chiral1} becomes
\bea
S \, = \,  \, \int d^4 x \sqrt{-g}  \, f(\phi) \, \varepsilon^{\mu\nu\a\b} R_{\a\b\r\s} 
\left( R_{\m\lambda}{}^{\r\s} \phi_\n \phi^\lambda  - \frac{1}{4} X R_{\m\n}{}^{\r\s}\right) \, .\label{chiralex}
\eea
Thus, by construction, we get the primary constraints of the form $\chi^{ij}  =  P^{ij}  - \Pi^{ij} \approx 0$ with
\bea
\Pi^{ij} = &-& 2 f(\phi) \left\{ \phi_m \phi^m \left( \e^{ik\l}D_\l Q^j_k + \e^{jk\l}D_\l Q^i_k \right) - \phi_k \e^{k\l m} \left( \phi^j D_m Q^i_\l + \phi^i D_m Q^j_\l \right) \right. \nb \\
&+& \left. \phi^k \phi_\l \left[ \e^{j\l m} \left( D_k Q^i_m - D_m Q^i_k \right) + \e^{i\l m} \left( D_k Q^j_m - D_m Q^j_k \right) \right] \right\} \, .\label{Pi1}
\eea
Notice that $\Pi^{ij}$ in the above expression turns out to be traceless and it is therefore useful to decompose these six primary constraints into trace and traceless parts,
\bea\label{PPi}
P \approx 0 \,, \qquad \tilde \chi^{ij}  =  \tilde P^{ij}  - \tilde \Pi^{ij} \approx 0 \,,
\eea
where we used the same notations as in \eqref{primaryP}.

Remarkably, the tuning (\ref{cond1}) not only leaves the action linear in $\dot Q_{ij}$, but also in $\dot \phi$.
Therefore, we get one additional primary constraint:
\beq
\pi_\phi - \varphi \approx 0 \,, \qquad \varphi \equiv \frac{\partial S}{\partial \dot \phi} \,,
\eeq
where $S$ is the action (\ref{chiralex}). In other words, 
$\varphi$ contains all the terms proportional to $\dot \phi$ in the action and its explicit form is not needed.

Using the expression (\ref{Pi1}) one can compute the Dirac matrix $\tilde \Delta^{ij,k\l}(x,y)= \{\tilde \chi^{ij}(x), \tilde \chi^{k\l}(y)\}$ between the primary constraints $\tilde \chi^{ij}$ and find out that it is not completely degenerate.
This means that not all the 5 primary constraints $\tilde \chi^{ij}$, associated to the higher derivative modes in the Lagrangian, lead to secondary constraints. Hence, the action (\ref{chiralex}) is expected to contain \Ost modes.

In this case, even considering the restriction to the unitary gauge, where the scalar field is assumed to depend  on time only, does not help since the action (\ref{chiralex})  identically vanishes. However, one can go back  to the full action (\ref{chiral1}) and study it in the unitary gauge, as we do just below.

\subsubsection{Unitary gauge}

\no Let us analyse the action (\ref{chiral1}) in the unitary gauge. We express it in a first order formulation, and we still get primary
constraints of the form \eqref{PPi} with now 
the following expression for the (traceless) $\Pi^{ij}$ tensor
\beq
\Pi^{ij}=\tilde \Pi^{ij} = - \left( 4 a_1 + 2 a_2 + a_3 + 8 a_4 \right) \frac{\dot \phi^2}{N^2} \left( \e^{ik\l}D_\l Q^j_k + \e^{jk\l}D_\l Q^i_k \right) \,. \label{Pi1UG}
\eeq
In contrast with the CS term in the unitary gauge, the five primary constraints $\tilde \chi^{ij}  \approx 0$ do not generate 5 secondary constraints, because of the presence of the lapse function in the denominator of (\ref{Pi1UG}).

As a consequence, the only way out is to tune the functions $a_A$ in order to eliminate the $\Pi^{ij}$ tensor~(\ref{Pi1UG}) itself, namely removing any higher order derivative in the action. This requirement gives the condition
\beq
4 a_1 + 2 a_2 + a_3 + 8 a_4 = 0 \,. \label{cond1UG}
\eeq
Solving for instance for the function $a_3$, we obtain the following action
\bea
S_{\rm UG} &=& \frac{2 \, \dot \phi^2 \, \e^{ij\l}}{N} \left[ 2 \left( 2 a_1 + a_2 + 4 a_4 \right) \left( K K_{mi} D_\l K^m_j 
+  {}^{(3)}\!R_{mi} D_\l K^m_j -  K_{mi} K^{mn} D_\l K_{jn} \right) \right.\nb \\
&-& \left. \left( a_2 + 4 a_4 \right) \left( 2\,  K_{mi} K^n_j D_n K^m_\l +  {}^{(3)}\!R_{j\l m}{}^n D_n K^m_i
\right)
\right] \,. \label{chiral1UG}
\eea

The action (\ref{chiral1UG}) does not involve any higher order time derivative of the metric and in this form represent a parity breaking extension of Horava--Lifshitz gravity. It is indeed clear that it propagates $[20 - (6 \times 2) - 2]/2 = 3$ DOF, exactly as does the Einstein-Hilbert action augmented with the CS term in the unitary gauge. However, its phenomenology should be completely different since, in the action (\ref{chiral1UG}), we do not have any higher order time derivative of the metric, but only higher order space derivatives.

\subsection{Including second derivatives of the scalar field}

\no In this final part, we enlarge the scope of our exploration to include theories with second derivatives of the scalar field in the action.
There is only a single Lagrangian that is linear in both the Riemann tensor and the second derivative of the scalar field $\phi$, namely
\beq
L_1 = \varepsilon^{\mu\nu\a\b} R_{\a\b\r\s}  \phi^\r \phi_\m \phi^\s_\n \,. \label{CL1}
\eeq
At the next level, i.e. still linear in the Riemann tensor but quadratically in the second derivative of $\phi$ (up to quadratic order in first derivatives of $\phi$), we find 6 independent Lagrangians:
\bea
&& L_2 = \varepsilon^{\mu\nu\a\b} R_{\a\b\r\s} \phi_\m^\r \phi^\s_\n \,, \qquad\qquad
L_3 = \varepsilon^{\mu\nu\a\b} R_{\a\b\r\s}  \phi^\s  \phi_\m^\r \phi^\lambda_\n \phi_\lambda \,, \nb \\
&& L_4 = \varepsilon^{\mu\nu\a\b} R_{\a\b\r\s}  \phi_\n  \phi_\m^\r \phi_\lambda^\s \phi^\lambda \,, \qquad
L_5 = \varepsilon^{\mu\nu\a\b} R_{\a\r\s\lambda}  \phi^\r \phi_\b \phi_\m^\s \phi^\lambda_\n \,, \nb \\
&& L_6 = \varepsilon^{\mu\nu\a\b} R_{\b\g}  \phi_\a  \phi_\m^\g \phi^\lambda_\n \phi_\lambda \,, \qquad\quad
L_7 =  (\Box \phi) L_1 \,. \label{CL2}
\eea
We will not investigate  here terms of higher order and thus simply consider the general linear combination of the above terms
\beq
S = \sum_{A = 1}^7 \int d^4 x \sqrt{-g} \, b_A L_A \,, \label{chiral2}
\eeq
where $b_A(\phi, X)$ are functions of the scalar field and its kinetic term.

\subsubsection{Brief Hamiltonian analysis}

The action (\ref{chiral2}) now involves second time derivatives of the scalar field, 
 in addition to the second time derivative of the spatial metric $\g_{ij}$. As a consequence, it is useful to perform a first order reformulation of the action also for taking into account $\ddot \phi$, in the same way as we do for $\ddot \g_{ij}$ (see section~\ref{Csec}).
For this purpose, let us introduce a one-form $A_\m$ that will replace $\phi_\m$ in the Lagrangians (\ref{CL1})~--~(\ref{CL2}) and add to the action (\ref{chiral2}) the following constraint through a Lagrangian multiplier~\cite{Langlois:2015cwa}
\beq
A_\m -\phi_\m \; = \; 0 \,.
\eeq
The one-form $A_\m$ decomposes in its time ($A_*$) and spatial ($\hat A_\mu$) projections
\beq
A_\m = - A_* n_\m + \hat A_\m \,,
\eeq
and using the fact that $\nabla_\m A_\n = \nabla_\n A_\m$, we get the following ADM decomposition for the derivative of $A_\m$~\cite{Langlois:2015cwa}
\beq
\nabla_\m A_\n = \frac{n_\m n_\n}{N} \left( \dot A_* - {\cal L}_{\vec{N}} A_* - \hat A^\r D_\r N  \right) - A_* K_{\m\n} + 2\, n_{(\m} K_{\n)\r} \hat A^\r - 2\, n_{(\m} D_{\n)} A_* + D_{(\m} \hat A_{\n)} \,.
\eeq
Substituting this decomposition to the action \eqref{chiral2}, one obtains the first order form that is needed to start the Hamiltonian 
analysis.

At this stage we have a priori 6 \Ost modes described by the $Q_{ij}$ variables and 1 additional  \Ost mode described by $A_*$. 
In order to get  rid of all of them, the generalization of the  Hessian matrix  (\ref{kinetic matrix}) that  includes $\dot A_*$ must be fully degenerate, which means that the action must not contain terms quadratic in $\dot Q_{ij}$ or $\dot A_*$. 
Because of the $\varepsilon$ tensor, the action is automatically devoid of  terms quadratic in $\dot A_*$, but it does contain mixed terms  $ \dot A_*\dot Q_{ij}$ in general. The latter disappear if one imposes the conditions
\beq
b_7 = 0 \qquad \text{and} \qquad b_6 = 2 \left( b_4 + b_5 \right) \,,\label{cond2}
\eeq
which we now assume. 
In this case, one gets  7 primary constraints of the form
\beq
\pi_{A_*} - \a \approx 0 \,, \qquad \chi^{ij}  = P^{ij}  - \Pi^{ij} \approx 0 \,,
\eeq
where $\a$ contains all the terms proportional to $\dot A_*$ in the action.

The Dirac matrix $\Delta^{ij,k\l}(x,y) = \{\chi^{ij}(x), \chi^{k\l}(y)\}$ between the constraints $\chi^{ij}$ turns out to be 
non-degenerate and no choice of functions, except the trivial one,  can make it vanish.
Therefore, the time evolution of the 6 primary constraints $\chi^{ij}$ determines the Lagrange multipliers $\xi_{ij}$ and no secondary constraint is generated.
In conclusion,  the action (\ref{chiral2}) with conditions (\ref{cond2}) contains 3 \Ost modes in the metric sector.

\subsubsection{Unitary gauge}
\no
Let us now examine the restriction to the unitary gauge of the action (\ref{chiral2})  with the conditions~(\ref{cond2}).
 In the unitary gauge, the scalar field depends only on time and therefore the components of $A_\m$ reduce to
\beq
A_* = \frac{\dot \phi (t)}{N} \,, \qquad \hat A_i = 0 \,,
\eeq
while the free functions $b_A$  depend now on $t$ and $N$ only.
The $\Pi^{ij}$ tensor becomes traceless and the Dirac matrix $\tilde \Delta^{ij,k\l}(x,y)$ simplifies to
\bea
\{\tilde \chi^{ij}(x), \tilde \chi^{k\l}(y)\} &=& - \frac{\dot \phi^2}{N^5}\left[ 2\, b_2 N^2 + \dot \phi^2 \left( b_3 - b_4 \right) \right]  \\
&& \left( \e^{ikm} \, \g^{j\l} + \e^{jkm} \, \g^{i\l} + \e^{i\l m} \, \g^{jk} + \e^{j\l m} \, \g^{ik} \right)  \partial_m N(x) \, \d (x-y) \,, \nb
\eea
which vanishes if
\beq
b_2 =  - \frac{A_*^2}{2} \left( b_3 - b_4 \right)  \,. \label{cond3}
\eeq
However, the above condition also removes  all 
the $\dot Q_{ij}$ terms in the action, which then reduces, in the unitary gauge, to
\bea
S_{\rm UG} &=& \frac{\dot \phi^3}{N^4} \, \e^{ij\l}  \left\{ 2\, N \left[ b_1 N K_{mi} D_\l K^m_j 
+ \left(b_4 + b_5 - b_3 \right) \dot \phi K_{mi} K^n_j D_n K^m_\l \right]  \right.\nb \\
&+& \left. \dot \phi  \left[ b_3 \;  {}^{(3)}\!R_{j\l m}{}^n K^m_i D_n N - 2 \left( b_4 + b_5 \right) \; {}^{(3)}\!R_{m\l} K^m_j D_i N \right]
\right\} \,. \label{chiral2UG}
\eea
The theory defined by (\ref{chiral2UG}) propagates only $[20 - (6 \times 2) - 2]/2 = 3$ DOF.  This is the same number of degrees of freedom as found for  (\ref{chiral1UG}) but one can note that  the present action involves also space derivatives of the lapse function, due to the higher order derivatives of the scalar field.

\bigskip

In principle, one could apply the same type of analysis for more complicated Lagrangians. Our results for the ``simplest'' Lagrangians do not lead us to believe that one would find a theory devoid of \Ost ghosts in its fully covariant version. 
So far, one can conclude from our exploration that the theories we already studied should be considered as low energy EFT or as Lorentz breaking ones,
on the same footing as Chern-Simons or Horava--Lifshitz gravity respectively.
In that respect, we leave the phenomenological study of both the actions (\ref{chiral1UG}) and (\ref{chiral2UG}) for future work.

\section{Conclusions}
\label{Conc}

\no In this paper, we have studied fully and partially degenerate metric theories in four dimensions
whose action is invariant under diffeomorphisms and contain at most second derivatives of the metric. Apart from the 
Einstein-Hilbert action which propagates two physical degrees of freedom, fully degenerate theories are either trivial 
(which correspond to the Gauss-Bonnet and the Pontryagyn Lagrangians) with no degree of freedom, 
or contain Ostrograsky modes (which is the case for the cubic $C$ Lagrangian). We have performed
a complete Hamiltonian analysis of the $C$ Lagrangian which shows that the theory indeed contains 5 DOF, 3 of them being 
Ostrogradsky ghosts, as confirmed by the analysis of linear perturbations.

We have also considered partially degenerate theories 
whose Lagrangian is given by an arbitrary (non-linear)  function  of  one of the fully degenerate Lagrangians, i.e.  $f(Y)$ Lagrangians, with $Y=R,GB,P$. More general partially degenerate Lagrangians (depending on several of the $Y$'s) are discussed 
in an appendix. Following the conservative criterion we set in the Introduction, i.e. that each (second class) primary constraint needs to generate a secondary constraint in order to remove the Ostrogradsky ghost,
we conclude that, apart from $f(R)$, partially degenerate theories seem to contain Ostrogradsky modes. 
$f(GB)$, after being reformulated as a scalar-tensor theory, can  easily be  cured by adding a kinetic term for the scalar field.
$f(P)$ instead, which can be reformulated as non-dynamical
Chern-Simons plus a potential, contains three extra modes, 
equally for the dynamical case.
However, when one restricts  Chern-Simons  modified gravity to the unitary gauge where the scalar field is a function of time only, one obtains a Lorentz breaking theory where all the \Ost modes are removed. 

Finally, we considered new parity breaking scalar-tensor theories constructed by combining  the Riemann tensor and the (first or second) derivatives of the scalar field.
Even though they contain \Ost modes in their covariant version, we have classified new classes of ``chiral scalar-tensor theories" which propagate only three degrees of freedom in the unitary gauge. 
In this sense, they have to be considered as generalizations of Chern-Simons modified gravity, i.e. as low energy EFTs, or as Lorentz breaking theories  with a  parity violating sector. 

Various phenomenological developments in these new theories are worth exploring: in particular the propagation of gravitational waves and black hole solutions.
A preliminary study shows that it is only when we introduce metrics that break parity, such as rotating axisymmetric geometries, that these terms kick in modifying GR solutions, while admitting certain GR solutions notably Schwarzschild in the other cases (see \cite{Campbell:1990fu} for similar behaviour in CS gravity).

\section*{Acknowledgments}
\no We thank Kazuya Koyama, Gianmassimo Tasinato, Alex Vikman and Nicol\'as Yunes for many interesting discussions and correspondence.
MC is supported by the European Research Council through grant 646702 (CosTesGrav). He is also grateful to LPT-Orsay for hospitality and financial support during the early stage of this work.
This work was supported by the Programme National Cosmology et Galaxies (PNCG) of CNRS/INSU with INP and IN2P3 and from DEFI InFIniTI, CNRS/INP.

\appendix

\section{Instability of linear perturbations for $C$}
\label{Cpert}
\no To illustrate the instability of the theory defined by \eqref{cubic action}, we make a linear perturbation analysis of the
theory and we show that the perturbations are indeed unstable.

Let us consider a background metric $\overline{g}_{\mu\nu}$, solution to the equations of motion, and a small 
perturbation $h_{\mu\nu}$ around this background. Plugging $g_{\mu\nu}=\overline{g}_{\mu\nu}+h_{\mu\nu}$ into the action leads 
generically to a quadratic action for the perturbation of the form
\bea
S_{quad}[h_{ij}] \; = \; \int dt\, d^3x \left\{ \ddot{h}_{ij} \left(A^{ijkl} h_{kl} + B^{ijkl} \dot{h}_{kl} + C^{ijklm} \overline{\nabla}_m h_{kl}+
D^{ijklm}\overline{\nabla}_m \dot{h}_{kl} \right)+ E \right\}
\eea
where the tensors $A$, $B$, $C$ and $D$ are evaluated in the background, $\overline{\nabla}$ denotes the covariant derivative
compatible with the background metric $\overline{g}_{\mu\nu}$ and we have included in E all terms which do not involve second time derivative of the perturbation. 

Integration by parts allows to simplify the term in the action which is linear in $\ddot{h}_{ij}$ as follows
\bea
S_{quad}[h_{ij}] \; = \; \int dt\, d^3x \left\{ \ddot{h}_{ij} \left( B^{ijkl} \dot{h}_{kl} + D^{ijklm}\overline{\nabla}_m \dot{h}_{kl} \right)+ E\right\}
\eea
with a redefinition of $E$.
Now, if we make a Fourier transform in the space coordinates, we find that the dynamics of the Fourier components ${\varphi}_{ij}$ 
of $h_{ij}$ is governed by an action of the type
\bea
\hat{S}_{quad}[\varphi_{ij}] \; = \; \int dt \, d^3k \left\{ \ddot{\varphi}_{ij} \, {\cal K}^{ijkl} \, \dot{\varphi}_{kl} + {\hat E}\right\}
\eea
where ${\cal K}^{ijkl} $ is evaluated in the background but could depend on wave number and $\hat E$ is the spatial Fourier transform of $E$. Only the skew symmetric component of $\cal K$ is relevant
for us because the symmetric component leads (after an integration by parts) to a term which involves only first time derivatives. Hence, without loss of generality, we assume that $\cal K$ is skew symmetric. It is well-known that any skew symmetric matrix can be brought to a block
diagonal form by a special orthogonal transformation. As $\cal K$ is a $6\times 6$ matrix, its block diagonal form is
\bea
\left(
\begin{array}{cccccc}
0 & \kappa_1 & 0 & 0 & 0 & 0 \\
-\kappa_1 & 0 & 0 &0 &0 & 0 \\
0 & 0 & 0 & \kappa_3 & 0 & 0 \\
0 & 0 & -\kappa_3 & 0 & 0 & 0 \\
0 & 0 & 0 & 0 & 0 & \kappa_5 \\
0 & 0 & 0 & 0 & -\kappa_5 & 0 \, ,
\end{array}
\right)
\eea
where $\kappa_A$ depend on the explicit form of $\cal K$. 
Therefore, a change of variable $(\varphi_{ij}) \mapsto (\varphi_1,\cdots,\varphi_6)$ 
allows us to decouple the different components of $h_{ij}$  in such a way that the action reduces to
\bea
 \int dt \, d^3k \left\{ \kappa_1 \, \ddot{\varphi_1} \dot{\varphi_2} + \kappa_3 \, \ddot{\varphi_3} \dot{\varphi_4} +\kappa_5\,  \ddot{\varphi_5} \dot{\varphi_6} )+ \hat E \right\}\, .
\eea
To study the stability of the perturbation, we  proceed as usual and we  replace this action by the following equivalent one
\bea
 \int dt \, d^3k \left\{ \kappa_1 \, \dot{\Phi}_1 \dot{\varphi_2} + \kappa_3 \, \dot{\Phi}_3 \dot{\varphi_4} +
 \kappa_5\,  \dot{\Phi}_5 \dot{\varphi_6} + \hat E + \pi_1(\dot{\varphi}_1 - \Phi_1) +  
 \pi_3(\dot{\varphi}_3 - \Phi_3)  + \pi_5(\dot{\varphi}_5 - \Phi_5) \right\}
\eea
which is clearly not degenerate (when $\kappa_A$ are not vanishing, which is generically the case). 
Hence, one expects that  it contains \Ost ghosts.

\section{Theories with kinetic matrix of higher rank}
\label{higherank}
\no In this Appendix we shortly discuss theories of gravity whose kinetic matrix \eqref{kinetic matrix} is degenerate with a rank higher than 1. 
One can suspect that  all partially degenerate Lagrangians for a metric (whose Hessian matrix has a rank $r \geq$1) are
of the form  
\bea\label{L(RGBPC)}
\int d^4x \, \sqrt{-g} \, {\cal L}(R,GB,P,C) \, ,
\eea
where the Lagrangian $\cal L$ is a function of the four fully degenerate Lagrangians \eqref{fully}. 
However, we have no formal proof for showing this.
We have studied in  details the case
where $\cal L$ is a function of a single variable in Sect. \ref{pdt}, which corresponds to $r=1$. 
Here we are going to discuss the case where $\cal L$ is a function of more than one variable, which corresponds to degenerate theories 
with a kinetic matrix whose rank is $r>1$. 

\subsection{Kinetic matrix}
To show that these theories \eqref{L(RGBPC)} are indeed degenerate, let us compute the rank of its kinetic matrix.
First, we recall that each fully 
degenerate Lagrangian (denoted generically $X$) can be written in the form
\bea
\sqrt{-g} \, X \; = \; \dot{K}_{ij} \Pi_X^{ij} - V_X \, ,
\eea
where the explicit form of $\Pi_X^{ij}$ has been given for GB \eqref{PiGB} and P \eqref{PiP} only. The other two expressions can easily be
obtained but their explicit form are not needed. 
Hence, the kinetic matrix is easily computed and gives
\bea\label{kineticX}
{\cal A}^{ij, k\l}(x,y) \; = \; \left[ \sum_{X,Y}  {\mathbb H}^{X Y} \, \Pi_X^{ij}(x)\, \Pi_Y^{kl} (x)\right] \delta^3(x-y)\, , \quad
{\mathbb H}^{X Y} \equiv \frac{\partial^2 {\cal L}}{\partial X \partial Y}  \, ,
\eea
where the sum runs over the set $\{R,GB,P,C\}$ of all degenerate Lagrangians,
and $\mathbb H$ is a real and symmetric four-dimensional matrix.
Thus, $\mathbb H$ can be diagonalised according to
\bea
\mathbb H \; = \; \Lambda \, \left[\text{diag}(\lambda_1,\lambda_2,\lambda_3,\lambda_4) \right]\, {}^T\! \Lambda \, ,
\eea
where $\Lambda$ is orthogonal and $\text{diag}(\lambda_1,\lambda_2,\lambda_3,\lambda_4)$ is the diagonal matrix
of eigenvalues $\lambda_A$. As a consequence, the kinetic matrix \eqref{kineticX} can be viewed as a quadratic form which can be 
diagonalised itself as follows
\bea
{\cal A}^{ij, k\l} \; = \; \sum_{A=1}^4 \lambda_A \, {\Gamma}_A^{ij} {\Gamma}_A^{k\l} \, , \qquad
{\Gamma}_A^{ij} \equiv \sum_X \mathbb H^{AX} \Pi_X^{ij} \, .
\eea
For simplicity, we omitted to write the explicit space dependency.
Thus, when $\mathbb H$ is invertible, the kernel of ${\cal A}^{ij, k\l}$ is of dimension 2 corresponding to the two directions
orthogonal to the four (6-dimensional) vectors $\Gamma_A$: in that case, $\text{rank}({\cal A})=4$. 
In general, it is easy to see that
\bea\label{rank}
\text{rank}({\cal A}) \; = \; 4 - \text{corank}(\mathbb H) \; = \; \text{rank}(\mathbb H) \, ,
\eea
and, as expected, the theory \eqref{L(RGBPC)} is always (partially) degenerate. 
In particular, we recover the result of the previous subsection, 
namely $\text{rank}({\cal A})=1$ when ${\cal L}$ is non-linear function of a single variable.

\subsection{Constraint analysis}
To count the number of DOF, we make a Hamiltonian analysis. We proceed as in Sect. \ref{pdt}, and
we  replace \eqref{L(RGBPC)} by the equivalent first order action
\bea
\int d^4x \, \sqrt{-g} \left\{ {\cal L}(\dot{Q}_{ij} \Pi_X^{ij} - V_X) + 2N p^{ij}(K_{ij} - Q_{ij})\right\} \, ,
\eea
whose Lagrangian does not involve anymore second derivatives of the metric. 
The corresponding phase space is defined by the same Poisson structure as in \eqref{phase space}. Computing the momenta
$P^{ij}$ gives
\bea
P^{ij} \; = \; \sum_X  {\cal L}_X \, \Pi^{ij}_X \, , \qquad
{\cal L}_X\equiv \frac{\partial {\cal L}}{\partial X}  \, .\label{Pgen}
\eea
This relation immediately shows that the momenta are constrained. Indeed, when $f_X\neq 0$, one can use four out of the 
six equations \eqref{Pgen} to express the four functions $f_X$ in terms of the phase space variables, and the remaining
two equations are constraints. If only $r$ functions $f_X$ are different from zero, we only need $r$ equations to solve them
in terms of the phase space variables, and we get $c=6-r$ constraints denoted $\chi_c \approx 0$. 
This result is compatible with the formula \eqref{rank} as
$\text{rank}({\cal A})=6-c=r$ and $\text{rank}(\mathbb H)=r$. 

Hence, we start with $c$ primary constraints $\chi_c \approx 0$ in addition to the usual 4 constraints $\pi_\mu \approx 0$. The latter
lead to the usual 4  Hamiltonian and vectorial constraints which form with $\pi_\mu \approx 0$ a set of first class constraints (up to the
addition of second class constraints). The analysis of the stability under time evolution of  the constraints $\chi_c \approx 0$ 
is subtler. Even though we do not perform a complete analysis (which goes beyond the scope of this paper) here, let us quickly 
describe generic  cases.

When ${\cal L}$ is a function of $R$ and $GB$ only, one obtains $c$  secondary constraints, and there are no tertiary constraints. 
Except if ${\cal L}$ depends on $P$ only (in which case the theory admits a conformal invariance in addition to the diff invariance, as it was
shown in the previous section), all these $(2 \times c)$ constraints are second class. Thus, 
the theory propagates $[32-(2 \times 8) - (2\times c)]/2= (8-c)$ DOF. We recover the fact that $f(R)$ and $f(GB)$ propagates 3 DOF, whereas
$f(R,GB)$ propagates 4 DOF. See \cite{Comelli:2005tn} for an interesting class of $f(R,GB)$ theory.

When the Lagrangian ${\cal L}$ depends on $P$ and/or $C$, the analysis is more complicated. The time  evolution of the $c$ primary constraints does not produce generically secondary constraints. However, this is true only when $c$ is even, in which case the theory
propagates $[32-(2 \times 8) - c]/2= (8-c/2)$ DOF. When $c$ is odd, the Dirac matrix of the primary constraints is necessarily degenerate,
which implies that there is (generically) 1 secondary constraints. In that case, the theory propagates $(8-(c+1)/2)$ DOF. 
There might also be particular cases where there are more than 1 second class constraints. 
We leave a precise Hamiltonian analysis of such theories for the future.


\begin{thebibliography}{99}


\bibitem{Lovelock:1971yv} 
  D.~Lovelock,
  J.\ Math.\ Phys.\  {\bf 12}, 498 (1971).
  
\bibitem{Ostrogradski} 
M. Ostrogradsky. Mem. Ac. St. Petersbourg VI 4 (1850) 385;
  
  
\bibitem{Woodard:2006nt} 
  R.~P.~Woodard,
  Lect.\ Notes Phys.\  {\bf 720}, 403 (2007)
  [astro-ph/0601672].
  

\bibitem{Zumalacarregui:2013pma}
 M.~Zumalac�rregui and J.~Garc�a-Bellido,
 Phys.\ Rev.\ D {\bf 89} (2014) 064046
 [arXiv:1308.4685 [gr-qc]].

\bibitem{Gleyzes:2014dya}
 J.~Gleyzes, D.~Langlois, F.~Piazza and F.~Vernizzi,
 Phys.\ Rev.\ Lett.\  {\bf 114} (2015) no.21,  211101
 [arXiv:1404.6495 [hep-th]].
 
\bibitem{Gleyzes:2014qga}
 J.~Gleyzes, D.~Langlois, F.~Piazza and F.~Vernizzi,
 JCAP {\bf 1502} (2015) 018
 [arXiv:1408.1952 [astro-ph.CO]].
 
\bibitem{Lin:2014jga} 
  C.~Lin, S.~Mukohyama, R.~Namba and R.~Saitou,
  JCAP {\bf 1410}, no. 10, 071 (2014)
  [arXiv:1408.0670 [hep-th]].

\bibitem{Deffayet:2015qwa}
 C.~Deffayet, G.~Esposito-Farese and D.~A.~Steer,
 Phys.\ Rev.\ D {\bf 92} (2015) 084013
 [arXiv:1506.01974 [gr-qc]].
 
\bibitem{Crisostomi:2016tcp}
 M.~Crisostomi, M.~Hull, K.~Koyama and G.~Tasinato,
 JCAP {\bf 1603} (2016) no.03,  038
 [arXiv:1601.04658 [hep-th]].
 

\bibitem{Langlois:2015cwa}
 D.~Langlois and K.~Noui,
 JCAP {\bf 1602} (2016) no.02,  034
 [arXiv:1510.06930 [gr-qc]].
 
 \bibitem{Langlois:2015skt}
 D.~Langlois and K.~Noui,
 JCAP {\bf 1607} (2016) no.07,  016
 [arXiv:1512.06820 [gr-qc]].

 
 \bibitem{Crisostomi:2017aim} 
  M.~Crisostomi, R.~Klein and D.~Roest,
  JHEP {\bf 1706}, 124 (2017)
  [arXiv:1703.01623 [hep-th]].
 
 
 \bibitem{Chamseddine:2013kea}
  A.~H.~Chamseddine and V.~Mukhanov,
  JHEP {\bf 1311} (2013) 135
  [arXiv:1308.5410 [astro-ph.CO]].
 
 
\bibitem{Crisostomi:2016czh}
 M.~Crisostomi, K.~Koyama and G.~Tasinato,
 JCAP {\bf 1604} (2016) no.04,  044
 [arXiv:1602.03119 [hep-th]].
 
 \bibitem{Achour:2016rkg} 
  J.~Ben Achour, D.~Langlois and K.~Noui,
  Phys.\ Rev.\ D {\bf 93}, no. 12, 124005 (2016)
  [arXiv:1602.08398 [gr-qc]].
  
\bibitem{deRham:2016wji} 
  C.~de Rham and A.~Matas,
  JCAP {\bf 1606}, no. 06, 041 (2016)
  [arXiv:1604.08638 [hep-th]].
 
\bibitem{BenAchour:2016fzp} 
  J.~Ben Achour, M.~Crisostomi, K.~Koyama, D.~Langlois, K.~Noui and G.~Tasinato,
  JHEP {\bf 1612}, 100 (2016)
  [arXiv:1608.08135 [hep-th]].
  
\bibitem{Gleyzes:2014rba} 
  J.~Gleyzes, D.~Langlois and F.~Vernizzi,
  Int.\ J.\ Mod.\ Phys.\ D {\bf 23}, no. 13, 1443010 (2015)
  [arXiv:1411.3712 [hep-th]].

\bibitem{Langlois:2017mxy} 
  D.~Langlois, M.~Mancarella, K.~Noui and F.~Vernizzi,
  JCAP {\bf 1705}, no. 05, 033 (2017)
  [arXiv:1703.03797 [hep-th]].
  
\bibitem{Heisenberg:2016eld} 
  L.~Heisenberg, R.~Kase and S.~Tsujikawa,
  Phys.\ Lett.\ B {\bf 760}, 617 (2016)
  [arXiv:1605.05565 [hep-th]].
  
\bibitem{Kimura:2016rzw} 
  R.~Kimura, A.~Naruko and D.~Yoshida,
  JCAP {\bf 1701}, no. 01, 002 (2017)
  [arXiv:1608.07066 [gr-qc]].
  

\bibitem{Motohashi:2016ftl}
  H.~Motohashi, K.~Noui, T.~Suyama, M.~Yamaguchi and D.~Langlois,
  JCAP {\bf 1607} (2016) no.07,  033
  [arXiv:1603.09355 [hep-th]].
  
  
  \bibitem{Klein:2016aiq}
  R.~Klein and D.~Roest,
  JHEP {\bf 1607} (2016) 130
  [arXiv:1604.01719 [hep-th]].
  
\bibitem{Chen:2012au} 
  T.~j.~Chen, M.~Fasiello, E.~A.~Lim and A.~J.~Tolley,
  JCAP {\bf 1302}, 042 (2013)
  [arXiv:1209.0583 [hep-th]].
  
  \bibitem{Jackiw:2003pm} 
  R.~Jackiw and S.~Y.~Pi,
  Phys.\ Rev.\ D {\bf 68}, 104012 (2003)
  [gr-qc/0308071].  
  
\bibitem{Thomas} 
  T.~Y.~Thomas,
  ``The differential invariants of generalized spaces'',
  Cambridge University Press (1934).
  
\bibitem{Horndeski:2017rtl} 
  G.~W.~Horndeski,
  arXiv:1706.04827 [gr-qc].

\bibitem{Deruelle}
N. Deruelle, M. Sasaki, Y. Sendouda and D. Yamauchi, 
Prog. Theor. Phys. 123 (2010) 
[arXiv:0908.0679[hep-th]].

\bibitem{Bonifacio:2015rea} 
  J.~Bonifacio, P.~G.~Ferreira and K.~Hinterbichler,
  Phys.\ Rev.\ D {\bf 91}, 125008 (2015)
  [arXiv:1501.03159 [hep-th]].
  
\bibitem{Lovelock}
D. Lovelock,
Archive for Rational Mechanics and Analysis 33 (1969)

\bibitem{Lanczos:1938sf} 
  C.~Lanczos,
  Annals Math.\  {\bf 39}, 842 (1938).
  
  \bibitem{Sotiriou:2008rp}
  T.~P.~Sotiriou and V.~Faraoni,
  Rev.\ Mod.\ Phys.\  {\bf 82} (2010) 451
  [arXiv:0805.1726 [gr-qc]].
  
  
  \bibitem{DeFelice:2010aj}
  A.~De Felice and S.~Tsujikawa,
  Living Rev.\ Rel.\  {\bf 13} (2010) 3
  [arXiv:1002.4928 [gr-qc]].
  
  
  
\bibitem{Kobayashi:2011nu} 
  T.~Kobayashi, M.~Yamaguchi and J.~Yokoyama,
  Prog.\ Theor.\ Phys.\  {\bf 126}, 511 (2011)
  [arXiv:1105.5723 [hep-th]].  
  
  
\bibitem{Alexander:2009tp} 
  S.~Alexander and N.~Yunes,
  Phys.\ Rept.\  {\bf 480}, 1 (2009)
  [arXiv:0907.2562 [hep-th]].  
  
\bibitem{Dyda:2012rj} 
  S.~Dyda, E.~E.~Flanagan and M.~Kamionkowski,
  Phys.\ Rev.\ D {\bf 86}, 124031 (2012)
  [arXiv:1208.4871 [gr-qc]].  
  
\bibitem{Delsate:2014hba} 
  T.~Delsate, D.~Hilditch and H.~Witek,
  Phys.\ Rev.\ D {\bf 91}, no. 2, 024027 (2015)
  [arXiv:1407.6727 [gr-qc]].
  
\bibitem{Okounkova:2017yby} 
  M.~Okounkova, L.~C.~Stein, M.~A.~Scheel and D.~A.~Hemberger,
  Phys.\ Rev.\ D {\bf 96}, no. 4, 044020 (2017)
  [arXiv:1705.07924 [gr-qc]].  
  
 \bibitem{Boulware}  
D. G. Boulware,
Quantization of higher derivative theories of gravity,
in {\it Quantum Theory of Gravity: Essays in Honor of the Sixties Birthday of Bryce S. DeWitt},
edited by S. M. Christensen (Adam Hilger, Bristol, England, 1984), pp. 267--294.
 
\bibitem{Campbell:1990fu} 
  B.~A.~Campbell, M.~J.~Duncan, N.~Kaloper and K.~A.~Olive,
  Nucl.\ Phys.\ B {\bf 351}, 778 (1991).
  D.~Grumiller and N.~Yunes,
  Phys.\ Rev.\ D {\bf 77}, 044015 (2008)
  [arXiv:0711.1868 [gr-qc]].
   
\bibitem{Comelli:2005tn} 
  D.~Comelli,
  Phys.\ Rev.\ D {\bf 72}, 064018 (2005)
  [gr-qc/0505088].
  
  
  
  
  




\end{thebibliography}
\end{document}